\documentclass[american]{article}
\usepackage[T1]{fontenc}
\usepackage[latin9]{inputenc}
\usepackage{units}
\usepackage{amsmath}
\usepackage{amssymb}
\usepackage{graphicx}

\makeatletter

\providecommand{\tabularnewline}{\\}

\usepackage[left=3cm,right=2cm,includeheadfoot]{geometry}
\makeatother

\usepackage{babel}
\begin{document}
\title{Refraction of light and conservation laws}
\author{R. Dengler \thanks{ORCID: 0000-0001-6706-8550}}
\maketitle
\begin{abstract}
The refraction of light by dispersion-free dielectric media can be
modeled using well-localized macroscopic wave packets, enabling a
description in terms of pseudo-particles. This approach is often used
in thought experiments to illustrate aspects of the Abraham-Minkowski
debate. This work uses the particle picture to show at an elementary
level how different types of momenta come into play, and how light
refraction can be explained at the level of particles. A special exactly
solvable microscopic model is used to illustrate the interplay and
tension between microscopic physics and the conventional effective-medium
Maxwell equations.
\end{abstract}

\section{Introduction}

\begin{figure}
\centering{}\includegraphics[scale=1.2]{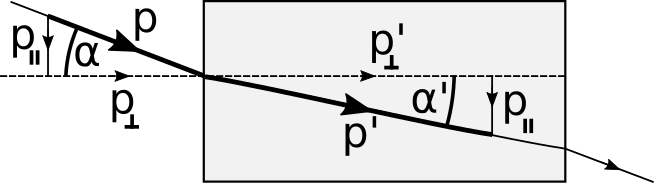}\caption{\protect\label{fig:Refraction}Law of refraction in the corpuscular
theory of light. Conservation of the parallel component of momentum
($p_{\Vert}$) at the surface of incidence, assuming constant total
momentum $p$ outside and $p'$ inside the medium, leads to the law
of sines.}
\end{figure}

In Newton's corpuscular theory of light, light is composed of of tiny,
massive corpuscles, and the sine law of refraction is explained using
a mechanical model (proposition 95 of book 1 of his Principia (1686)).
In modern terms, Newton proposed that a refracting body acts as 
a potential well of constant depth for the corpuscles, causing a higher
speed within the medium due to energy conservation. Upon entering
the medium the perpendicular component of momentum changes, while
the component $p_{\Vert}$ parallel to the surface remains unchanged.
This relationship is expressed as
\begin{equation}
p_{\Vert}=p\sin\alpha=p'\sin\alpha',\label{eq:SnellNewton}
\end{equation}
where $p$ is total momentum in vacuum, $p'$ total momentum in the
medium, and $\alpha$ and $\alpha'$ are the angle of incidence measured
from the normal vector, see Fig.~\ref{fig:Refraction}. If $p$ is
constant then $p'$ only depends on the type of medium, and also is
constant, and equation (\ref{eq:SnellNewton}) is Snell's law, the
law of sines.

The corpuscular theory of light gradually lost its status in favor
of the wave theory of light, largely due to the work of C. Huygens,
T. Young, and A. J. Fresnel. The possibility to measure wavelengths
also demonstrated that light propagates in matter with a speed reduced
by a factor $n$, the refractive index, while Newton's model requires
greater speed in the medium, in contradiction also to special relativity.
Nevertheless, the concept of light particles reemerged in the form
of photons within the framework of quantum field theory.

Newton's original model cannot be reconciled with modern physics.
However, by considering the refraction of localized wave packets with
definite momentum and energy, it should be possible to explain the
kinematics of refraction in terms of particles, wave packets of light
in vacuum and pseudo-particles \footnote{One could also speak of photons in the vacuum and quasiparticles in
the medium, but what is meant here are macroscopic classical signals,
which can be measured and observed along their path like classical
particles.} in the medium.

This would again represent a valid mechanistic  model for refraction. 

The basic assumptions for an idealized physical scenario can be summarized
as follows:
\begin{itemize}
\item \label{enu:Large}Wave packets are small, with a narrow frequency
range, system can be large.
\item \label{enu:Dispersion}Dispersion can be ignored.
\item \label{enu:Tranverse}Transverse polarization does not play a role.
\item \label{enu:LocalInt}Interactions are local.
\item \label{enu:QM_Avg}Quantum randomness does not play a role (surfaces
can have an anti-reflection coating). 
\item \label{enu:Electrostrict}Electrostriction is negligible.
\item \label{enu:Dissipation}Dissipation is negligible.
\item \label{enu:DirMomentum}Refraction does not transfer momentum in directions
parallel to the surface.
\end{itemize}
The last condition is not as intuitive or self-evident as it might
seem. Translational invariance in the direction parallel to the surface
requires only conservation of the component of total momentum in this
direction. A counterexample is a charged particle in a continuous
resistive medium.

The essential additional input is that the wavelength $\lambda$ in
a medium with refractive index $n$ is reduced by a factor $1/n$.
Correspondingly, the signal speed becomes $c/n$, where $c$, is the
speed of light in a vacuum. 

\section{Conservation laws}

Under these assumptions, wave packets behave like macroscopic particles
with localized energy and momentum, and therefore must satisfy the
standard conservation laws. 

\subsection{Momentum conservation}

If the component of signal momentum parallel to the surface is conserved,
and if $p$ represents signal momentum in vacuum, then Snell's law
implies that signal momentum in the medium is given by 
\begin{equation}
p'=pn.\label{eq:p_Minkowski}
\end{equation}
Newton explained this by conceiving the medium as a potential well,
but this contradicts the lower signal speed in the medium. Instead,
the signal in the medium must be interpreted as a different type of
particle. The momentum (\ref{eq:p_Minkowski}) for a signal in a medium
also is referred to as \emph{Minkowski momentum} \cite{BulgeExp2022,Barn2010,BarnLoud2010,Mansu2010}.

An argument in favor of Minkowski momentum (\ref{eq:p_Minkowski})
is that it agrees with the quantum mechanical formula for the momentum
$p_{QM}=\hbar k$ of a quasiparticle, where $\hbar$ is Planck's constant
and $k=2\pi/\lambda$ is the wavevector. The momentum (\ref{eq:p_Minkowski})
also appears to agree with measurements in liquids or gases, for a
review see~\cite{Kemp2011}.

Conservation of momentum implies that a momentum $\pm\left(p-p'\right)$
is transferred to the surface of the medium when the signal enters
or exits \footnote{In the natural scenario, without an anti-reflective coating, the \emph{additional}
incoming and reflected the signal also transfers momentum to the surface.}. This sudden momentum transfer generates small sound waves propagating
into the medium. It is not difficult to verify that these sound waves
contain a negligible amount of energy in realistic situations (section
\ref{sec:Dissip}). A superficial analysis might end here, but there
is a caveat, and this is also where the Abraham-Minkowski controversy
comes into play. 

\subsubsection{Connection to Abraham-Minkowski debate}

For many scenarios propagation of light in a medium is well understood.
If the wavelength is much larger than the distance between atoms,
Maxwell's equations combined with the linear relation $D=\epsilon E$
between the electric displacement field $D$ and the electric field
$E$, are sufficient to derive the rules of optics. This includes
the laws of reflection and refraction for different polarizations,
as well as the intensities of reflected and transmitted light in both
isotropic and anisotropic media \cite{Jackson1975,Lipson1981}. Complications
include nonlinear dispersion, electro- and magnetostriction \cite{Partanen_2022,BulgeExp2022,Partanen_2023},
which also transfer momentum to matter.

However, even the fundamental problem of disentangling electromagnetic
and mechanical momentum is full of pitfalls. The root of the Abraham-Minkowski
controversy are the expressions of Minkowski and Abraham for electromagnetic
momentum in a medium, in terms of the electromagnetic fields $D$
and $B$ or $E$ and $H$. The expression should contain or should
not contain the material momentum, and one source of the difficulty
are the phenomenological equations $D=\epsilon E$ and $H=B/\mu$.
More recent reviews are \cite{Partanen_2022,Kemp2011}.

Yet, even after more than a century, the situation remains unsatisfactory.
A more fundamental question concerns the validity~\cite{Silveir2017}
of the usual Maxwell equations in matter, which replace atomic matter
with an averaged (coarse-grained) substance. Such an approximation
is adequate for effects depending on average polarizability and magnetizability.
Difficulties arise when attempting to calculate bilinear local microscopic
quantities, such as the momentum of matter. This issue is examined
in detail using a microscopic model in section~\ref{sec:LayerModel}.
Before that, however, we continue with the conservation laws.

\subsection{Center of energy}

The center of energy of a system is a rarely used concept of special
relativity, which enters here in two ways. The definition reads
\begin{equation}
\boldsymbol{X}\left(t\right)=\frac{\int\mathrm{d}^{d}xe\left(x,t\right)\boldsymbol{x}}{\int\mathrm{d}^{d}xe\left(x,t\right)},\label{eq:CentEnergy}
\end{equation}
where $e\left(x\right)$ denotes relativistic energy density. The
classical limit of $\boldsymbol{X}\left(t\right)$ is the center of
mass. 

One point is that any system (a wave packet, for example) behaves
as though its entire energy is concentrated at its center of energy,
at least with respect to translational degrees of freedom. More important
here is the fact that $\boldsymbol{X}\left(t\right)$ of an isolated
system is a linear function of time. A general proof of this fact
makes use of the continuity equations for momentum and energy \cite{Moller1952},
but no such details are required here.

\subsubsection{Constant speed of center of energy}

\begin{figure}
\centering{}\includegraphics[scale=1.3]{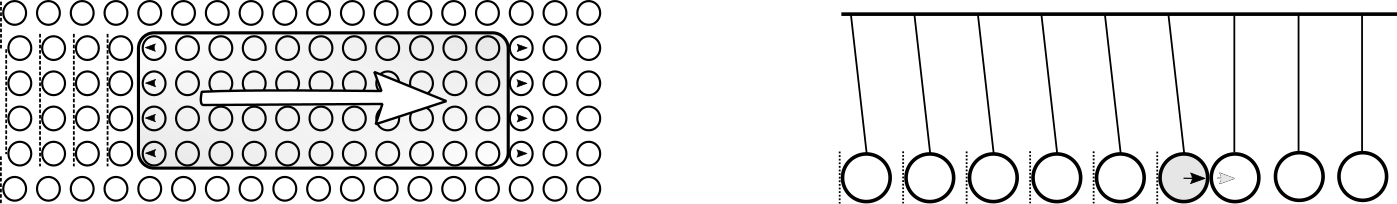}\caption{\protect\label{fig:Signal}Left: A scheme for a light signal in a
solid. The signal (the gray block) moves to the right, slightly accelerates
atoms at its right front to the right, and slightly decelerates atoms
at its left front, thus leaving behind a trace of atoms shifted to
the right, but again at rest.\protect \\
Right: A sequence of pendulums of elastic spheres as a metaphor for
a system where particle and signal speeds differ, and where the signal
leaves behind a trace of shifted spheres. Only the gray sphere is
in motion and on the verge of transferring its momentum to the next
sphere on the right.}
\end{figure}
The caveat mentioned above is the motion of the center of energy.
It is useful to now consider a slab of thickness $L$ and mass $M$
perpendicular to the $x$-axis, and a short wave packet of momentum
$p$ and energy $pc$ arriving from the left. To be able to compare
the incoming and the refracted signal we assume an anti-reflective
coating at both ends.

The signal traverses the slab in a time $t_{1}=Ln/c$, while a free
signal or photon would have reached position $Ln$ in this time. In
other words, refraction has shifted the contribution of the wave packet
with momentum $p$ to the center of energy by 
\begin{equation}
\Delta X_{\gamma}=\left(1-n\right)Lpc/\left(Mc^{2}+pc\right)\cong\left(1-n\right)Lp/\left(Mc\right)\label{eq:DeltaX_gamma}
\end{equation}
to the left. This must be compensated by a shift of the center of
energy (or mass) $\Delta X_{\mathrm{slab}}$ of the slab to the right.
The dilemma now is that the slab had momentum $p-pn$ during time
$t_{1}$, which gives 
\begin{equation}
\Delta X_{\mathrm{slab}}\cong\left(n-n^{2}\right)Lp/\left(Mc\right)<0,\label{eq:DeltaX_Slab}
\end{equation}
and thus $\Delta X_{\gamma}+\Delta X_{\mathrm{slab}}<0$, an apparent
contradiction.

The way out is to recognize that the Minkowski momentum (\ref{eq:p_Minkowski})\emph{
includes} a material part in such a way that the center of energy
$X$ moves with constant speed. This requires a contribution
\begin{equation}
\Delta X_{\mathrm{matter}}=-\Delta X_{\gamma}-\Delta X_{\mathrm{slab}}=\tfrac{Lp}{Mc}\left(n^{2}-1\right).\label{eq:DeltaX_M}
\end{equation}
It may seem confusing that there are now two material contributions
to $\Delta X$, i.e. $\Delta X_{\mathrm{slab}}$ and $\Delta X_{\mathrm{matter}}.$
However, $\Delta X_{\mathrm{matter}}$ arises from the signal itself,
and only involves a small fraction $m_{\mathrm{matter}}$ of the total
slab mass $M$, so it does not conflict with the expression for $\Delta X_{\mathrm{slab}}.$ 

One might argue that for the signal to have material momentum the
signal would have to carry some mass with itself with signal velocity
$c/n$, which is absurd in an insulator, where electrons and atoms
essentially stay at their place. Fig.~\ref{fig:Signal} shows how
the signal actually can be accompanied by a mass flow (similar illustrations
can be found in \cite{Mansu2010,Partanen_2017}). The signal slightly
accelerates the atoms at its front (on the right) and decelerates
the atoms on its rear. Atoms within the signal volume temporarily
are in motion, contributing both to momentum and mass flow. The signal
leaves behind a trace of atoms at rest, but slightly shifted to the
right. (The experiment involving a sequence of pendulums shown in
Fig.~\ref{fig:Signal} on the right is a metaphor for a signal leaving
behind a trace of shifted particles.) The momentum of the atoms within
the signal volume is
\begin{equation}
\Delta p_{\mathrm{matter}}=M\Delta X_{\mathrm{matter}}/t_{1}=p\left(n-\tfrac{1}{n}\right).\label{eq:Delta_p_Matt}
\end{equation}
This is Minkowski momentum (\ref{eq:p_Minkowski}) minus something
else, namely \emph{Abraham momentum} $p/n$. This suggests interpreting
Abraham momentum as the electromagnetic component of the signal's
momentum in the medium. As required $\Delta p_{\mathrm{matter}}>0$
overcompensates the momentum $p\left(1-n\right)<0$ of the slab as
a whole.

But how does this work for a continuous signal? In case of a continuous
signal, all transmitted photons are delayed by the slab, yet there
are no visible wave fronts that accelerate or decelerate the atoms.
The solution is that the initial wave front has accelerated the atoms
in its trace, and the momentum of these atoms, wherever it has gone,
compensates the delay of the photons. This continues until the terminating
wave front has passed through the medium. 

\subsubsection{Accelerating force}

The general picture displayed in Fig.~\ref{fig:Signal} was deduced
from conservation laws. Qualitatively, the force accelerating matter
can be explained by the longitudinal radiation pressure. This pressure
or stress is constant within signal. However, the pressure gradient
at the signal front generates a force that acts on the medium. Alternatively,
the force can be interpreted as the Lorentz force $j\times B$, acting
on the polarization current $j=\gamma$$\partial E/\partial t$, where
$\gamma$ represents polarizability and $E$ and $B$ are electric
and magnetic fields. This force averages to zero within the signal
, but not at the boundaries.

\section{Orders of magnitude, dissipation}

\label{sec:Dissip}The wave packet kinematics examined above is not
strictly reversible. The momentum mismatch $\pm\left(p'-p\right)$
at the boundaries generates small acoustic shock waves. Additionally,
the trace of shifted atoms left behind by the wave packet (Fig.~\ref{fig:Signal})
will relax with the speed of sound. These relaxation processes respect
conservation of momentum and the linear movement of center of energy,
and do not invalidate the general considerations. However, for consistency,
the order of magnitude of the implied dissipation should be evaluated.

\subsection{Momentum transfer at the boundaries}

At the boundaries it is natural to assume that the momentum $\pm\left(pn-p\right)$
is passed to a volume of approximately the same size as the signal
wave packet. If a significant part of the momentum $p=E/c$ of a light
signal with energy $E$ is transferred to a mass $m$ then the mass
acquires kinetic energy $T=\tfrac{1}{2}E^{2}/\left(mc^{2}\right)$.
Of interest is the ratio $E/\left(mc^{2}\right).$ 

For $E=1\unit{J}$ and $m=1\unit{mg}$ for a wave packet of volume
$1\unit{mm}^{3}$ in water one finds $E/\left(mc^{2}\right)\cong5\cdot10^{-12}.$
This means that only a minuscule fraction of the energy is lost at
the boundaries.

\subsection{Trace of shifted atoms}

The attenuation caused by the shifted atoms is of different nature.
The shift $\Delta X_{\mathrm{matter}}$ of the center of energy of
matter during the transmission of the signal from Eq.~(\ref{eq:DeltaX_M})
actually is caused by the shifted atoms left behind by the signal,
and not the entire slab mass $M$. If $m_{*}$ is the mass of the
atoms in the signal trace, then their shift 
\begin{equation}
\Delta X_{*}\sim\tfrac{Lp}{m_{*}c}\sim\tfrac{E}{\rho Ac^{2}}\label{eq:DeltaX_Column}
\end{equation}
is larger by a factor $M/m_{*}$. In the last expression we have used
$m_{*}=\rho AL$, where $\rho$ is mass density and $A$ the cross-sectional
area of the signal. For a signal in a cube of volume $1\unit{mm}^{3}$
with energy $E=1\unit{J}$ in water one finds the minuscule shift
$\Delta X_{*}\sim10^{-14}\unit{m}$.

The elastic energy of a cylinder in the slab of length $L$ shifted
by $\Delta X_{*}$ over a transverse distance $\sqrt{A}$ is
\begin{equation}
E_{\mathrm{elastic}}\sim G\sqrt{A}L\frac{\Delta X_{*}^{2}}{\sqrt{A}}\sim E\frac{v_{t}^{2}}{c^{2}}\tfrac{E}{\left(\rho c^{2}A^{3/2}\right)A^{1/2}}L\sim EL/\ell.\label{eq:E_Decay}
\end{equation}
Here $G$ is shear modulus, and $\sqrt{A}L$ the cylinder surface.
In the second expression the velocity $v_{t}=\sqrt{G/\rho}$ of transverse
sound and Eq.~(\ref{eq:DeltaX_Column}) were used. Eq.~(\ref{eq:E_Decay})
says that the signal decays exponentially with some characteristic
length $\ell$.  A sound velocity $v_{t}=1000\unitfrac{m}{s}$, a
signal energy $E=1\unit{J}$, a signal extension $\sqrt{A}=1\unit{mm}$
and the mass density $\rho$ of water give a decay length of $10^{19}\unit{m}$,
certainly negligible also in other realistic situations.

\section{An exactly solvable model}

\label{sec:LayerModel}
\begin{figure}
\begin{centering}
\includegraphics[scale=0.6]{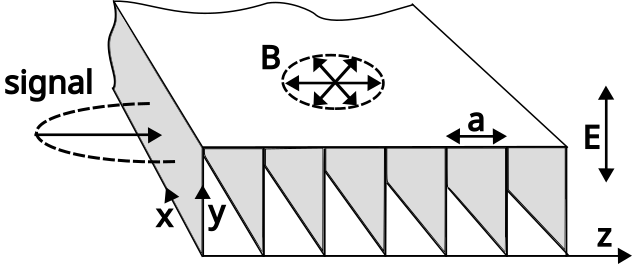}
\par\end{centering}
\centering{}\caption{\protect\label{fig:MetaMat}An optical metamaterial or a wave guide
consisting of thin polarizable layers perpendicular to the $z$-axis.
Signals propagate in  the $xz$-plane. The electric field $E$ always
points in $y$-direction, the magnetic field $B$ is in the $xz$-plane.}
\end{figure}

What still seems to be missing are exact solutions
for microscopic models. Here, we examine an optical metamaterial
consisting of thin polarizable layers oriented perpendicularly to
the $z$-axis at positions $z\in a\mathbb{Z}$, where $a$ is the
spacing, Fig.~\ref{fig:MetaMat}.
 We only consider signals propagating in the $x$-$z$ direction,
with the electric field oriented in the $y$ direction, parallel to
the layers \footnote{A similar three dimensional system has been investigated in~\cite{Draine_1993}
using discrete dipole approximation (DDA).}. There is no dependence on $y$. The medium could be restricted to
a finite $y$ range using two ideal conducting layers perpendicular
to the $y$-axis, without affecting the signals. The medium then would
be a planar capacitor or a planar wave guide parallel to the $xz$-plane.
For wavelengths much large than the layer spacing $a$ the signal
only sees the average capacitance per area rather than the individual
layers. In this limit the system thus effectively is a continuum,
isotropic in the $xz$-plane, and should reproduce the signal speed
and the law of refraction of the continuum model. This is indeed the
case, however, the internal signal properties differ.

The polarization density is $a\gamma E_{2}\sum_{m\in\mathbb{Z}}\delta\left(z-am\right)$,
the current density is
\begin{equation}
j_{2}=\gamma a\partial_{t}E_{2}\sum_{m}\delta\left(z-am\right),\label{eq:j3_exac}
\end{equation}
where $E_{2}$ is the electric field in $y$-direction and $\gamma$
denotes polarizability. The polarization implies a potential energy
$V=\tfrac{a}{2}\gamma E_{2}^{2}$ per layer area in the fictitious
oscillators in the layer material. The electrons in the material follow
the field in phase, their kinetic energy and momentum is negligible. 

One might object that infinitely thin layers with a finite polarizability
do not constitute a truly microscopic model. However, physical approximations
exist. One could envision a medium composed of atomic layers of a
dielectric material, introducing an atomic length scale. The granularity
affects the fields only in the immediate vicinity of the layers, within
several atomic scales. Polarization still is proportional to the electric
field near the layers, which is all that is required. This is a valid
physical realization of the model, exhibiting all relevant dependencies.
As shown below, knowing the fields between the layers suffices to
determine the mechanical momentum. To get a finite index of refraction,
the polarizability of the layers must be proportional to the layer
spacing. In theory this can be achieved by adjusting atomic polarizabilities.
The fact that this might be difficult with real atoms does not invalidate
the microscopic model.

For simplicity we now use $c=1$ for the speed of light. The Maxwell
equations in vacuum lead to an equation for $E_{2}$ alone,
\begin{align}
\partial_{x}^{2}E_{2}+\partial_{z}^{2}E_{2} & =\partial_{t}^{2}E_{2}+\gamma_{0}\partial_{t}^{2}E_{2}\sum_{m\in\mathbb{Z}}\delta\left(z-ma\right),\label{eq:EqMot_exac}\\
\gamma_{0} & =\gamma a/\epsilon_{0}=\left(n^{2}-1\right)a,\nonumber 
\end{align}
where $n$ is the refractive index. The law of induction $\partial_{t}B_{1}=\partial_{z}E_{2}$,
$\partial_{t}B_{3}=-\partial_{x}E_{2}$ allows to determine the magnetic
field. Together with the equations $\partial_{t}P_{m}=F_{m}$ for
the momenta $P_{m}$ of the layers according to the forces $F_{m}$
exerted on them, Eq.~(\ref{eq:EqMot_exac}) forms a closed set of
equations. The layers are assumed heavy, meaning they absorb momentum
and remain in place.

\subsection{Bloch solution}

The Bloch ansatz $E_{2}=e^{i\left(px+qz-\omega t\right)}\psi\left(z\right)$
with $\psi$ continuous and periodic in $a$ gives

\begin{align}
\left(\partial_{z}^{2}+2iq\partial_{z}-p^{2}-q^{2}+\omega^{2}\right)\psi & =-\gamma_{0}\omega^{2}\psi\sum_{m}\delta\left(z-ma\right),\label{eq:EqMotPsi}\\
\psi'\left(0^{+}\right)-\psi'\left(a^{-}\right) & =-\gamma_{0}\omega^{2}\psi\left(0\right),\nonumber 
\end{align}
where we have also written down the boundary condition at the layer.
In the interval $0\leq z<a$ the solution can be written in the form
\begin{equation}
\psi^{\left(0\right)}\left(z\right)=\alpha e^{-i\left(q-Q\right)z}+\beta e^{-i\left(q+Q\right)z}.\label{eq:psi_0}
\end{equation}
For wavelengths much larger than $a$ (linear dispersion) the solution
is

\begin{align}
\alpha & =\tfrac{1}{2}\left(1+q/Q\right),\label{eq:Obl_alpha_Q_om}\\
\beta & =\tfrac{1}{2}\left(1-q/Q\right),\nonumber \\
Q & =\tfrac{1}{n}\sqrt{q^{2}-\left(n^{2}-1\right)p^{2}},\nonumber \\
\omega & =\tfrac{1}{n}\sqrt{p^{2}+q^{2}}=\sqrt{p^{2}+Q^{2}}.\nonumber 
\end{align}
The phase and group speeds are $c/n$, as expected.

\subsection{Interpretation}

A peculiar property of the model is that the Bloch solution also describes
refraction. One can remove all layers at $z<0$ and extend the solution
(\ref{eq:psi_0}) from the interval $-a\leq z<0$ to the entire range
$z<0$ without changing anything else. This gives an incoming wave
with amplitude $\alpha$, a reflected wave with amplitude $\beta$
and the refracted $\alpha\beta$ signal in the medium at $z\geq0.$
The quantities $\left(p,Q\right)$ and $\left(p,q\right)$ are the
$x,z$ components of the wavevector in the vacuum and in the medium.
The refraction rate is the usual $\beta^{2}/\alpha^{2}$, for normal
incidence $\left(n-1\right)^{2}/\left(n+1\right)^{2}$ . 

What is unusual is that, on average, no force acts on the medium,
particularly on the first layer. This follows from the fact that all
layers see the same vacuum waves. (If one would add an additional
anti-reflective layer with appropriate polarizability at some position
$z<0$ then a force would act on this layer.)

It remains to calculate the energy densities $u$ and the momentum
densities $\Pi$. For $z\geq0$ this can be achieved with the fields

\begin{align}
E_{2} & =e^{i\left(px-\omega t\right)}\left(\alpha e^{iQz}+\beta e^{-iQz}\right)\cong e^{i\left(px-\omega t\right)},\label{eq:EB_in_Gap}\\
B_{1} & =e^{i\left(px-\omega t\right)}\left(-Q\tfrac{\alpha}{\omega}e^{iQz}+Q\tfrac{\beta}{\omega}e^{-iQz}\right)\cong-e^{i\left(px-\omega t\right)}q/\omega,\nonumber \\
B_{3} & =e^{i\left(px-\omega t\right)}\left(p\tfrac{\alpha}{\omega}e^{iQz}+p\tfrac{\beta}{\omega}e^{-iQz}\right)\cong e^{i\left(px-\omega t\right)}p/\omega.\nonumber 
\end{align}
in the first gap $0\leq z<a$. Actually $Qz$ only changes by a negligible
amount $Qa$ in the interval, and one can set $z=0.$ 

To simplify the comparison of the incoming and the refracted signal
it recommends itself to simulate an anti-reflective coating. This
corresponds to an incoming signal in the range $-\infty<z<0$ with
amplitudes $\alpha_{<}=\sqrt{\alpha^{2}-\beta^{2}}$ and $\beta_{<}=0$
instead of $\alpha$ and $\beta$, with the same energy flow (incoming
and reflected intensity $\beta^{2}$ removed). The results are listed
in table~\ref{tab:Densities}.

\begin{table}
\centering{}%
\begin{tabular}{|c|c|c|c|c|c|c|}
\hline 
\noalign{\vskip0.1cm}
 & $\left(k_{1},k_{3}\right)$ & $\overline{u^{\mathrm{vac}}}$ & $\overline{u^{\mathrm{tot}}}$ & $\overline{\Pi^{\mathrm{vac}}}$ & $\overline{\Pi^{\mathrm{mech}}}$ & $\overline{\Pi^{\mathrm{tot}}}$\tabularnewline
\hline 
\hline 
\noalign{\vskip0.1cm}
$z<0$ & $\left(p,Q\right)$ & $\tfrac{\epsilon_{0}}{2}\tfrac{q}{Q}$ & $\tfrac{\epsilon_{0}}{2}\tfrac{q}{Q}$ & $\tfrac{\epsilon_{0}}{2}\tfrac{q}{Q}$ & 0 & $\tfrac{\epsilon_{0}}{2}\tfrac{q}{Q}$\tabularnewline
\hline 
\noalign{\vskip0.1cm}
$z>0$ & $\left(p,q\right)$ & $\tfrac{\epsilon_{0}}{4}\left(n^{2}+1\right)$ & $\tfrac{\epsilon_{0}}{2}n^{2}$ & $\tfrac{\epsilon_{0}}{2}n$ & $\tfrac{\epsilon_{0}}{4}n\left(n^{2}-1\right)$ & $\tfrac{\epsilon_{0}}{4}n\left(n^{2}+1\right)$\tabularnewline
\hline 
\end{tabular}\caption{\protect\label{tab:Densities}Average energy densities $u$ and average
momentum densities $\Pi$. The latter are measured in the wavevector
direction, $\left(p,Q\right)$ or $\left(p,q\right)$, respectively.
Quantities for $z<0$ are for the incoming signal in the case of an
anti-reflective coating. The vacuum momentum density $\overline{\Pi^{\mathrm{vac}}}$
coincides with the Poynting vector (apart from a factor $c$), $\overline{\Pi^{\mathrm{tot}}}$
is the total momentum density. }
\end{table}
The overline denotes the time or space average calculated with the
real parts ($\Re$) of the fields. A useful formula is $\Re\left(a\right)\Re\left(b\right)=\tfrac{1}{2}\Re\left(ab+ab^{*}\right)$.
The condition for same energy flow at $z=0$ reads $\Pi_{<}^{\mathrm{vac}}Q=\Pi_{>}^{\mathrm{vac}}q/n$.
The dependence of the quantities for $z<0$ on the direction is an
artifact due to the normalization $\alpha+\beta=1$. A more practical
normalization has direction independent quantities for $z<0$ and
anisotropic quantities (a factor $Q/q$) in the medium.

\subsection{Mechanical momentum}

The energy densities $u$ and the Poynting vector $\Pi^{\mathrm{vac}}$
in vacuum immediately follow from the fields. The more interesting
quantity, the total momentum density in the medium, can been calculated
with the help of the stress tensor $\sigma_{ij}=\epsilon_{0}\left(-E_{i}E_{j}-B_{i}B_{j}+\tfrac{1}{2}\delta_{ij}\left(E^{2}+B^{2}\right)\right)$
in the gap $0\leq z<a$,  
\begin{align}
\overline{\sigma_{11}} & =\tfrac{\epsilon_{0}}{2}\left(\overline{B_{3}^{2}-B_{1}^{2}+E_{2}^{2}}\right)=\tfrac{\epsilon_{0}}{4}\left(\omega^{2}-q^{2}+p^{2}\right)/\omega^{2},\label{eq:sigma}\\
\overline{\sigma_{13}} & =\overline{\sigma_{31}}=-\epsilon_{0}\overline{B_{1}B_{3}}=\tfrac{\epsilon_{0}}{2}qp/\omega^{2},\nonumber \\
\overline{\sigma_{33}} & =\tfrac{\epsilon_{0}}{2}\left(\overline{B_{1}^{2}-B_{3}^{2}+E_{2}^{2}}\right)=\tfrac{\epsilon_{0}}{4}\left(\omega^{2}-p^{2}+q^{2}\right)/\omega^{2}.\nonumber 
\end{align}
The contraction with the unit vector $\hat{e}=\left(p/n\omega,q/n\omega\right)$
in propagation direction gives the momentum flow in the gap
\begin{equation}
\overline{\sigma_{i}^{\mathrm{tot}}}=\sum_{j}\overline{\sigma_{ij}}e_{j}=\tfrac{\epsilon_{0}}{4}\left(n^{2}+1\right)\hat{e}_{i}.\label{eq:momentum_flow}
\end{equation}
We now consider a plane wave with a wave front. Into an empty interval
of length $\ell$ (in a time $\ell n$) flows momentum $\tfrac{\epsilon_{0}}{4}\left(n^{2}+1\right)\ell n=\overline{\Pi_{>}^{\mathrm{tot}}}\ell$,
which gives the expression for total momentum density $\overline{\Pi_{>}^{\mathrm{tot}}}$.
However, the electromagnetic momentum in this interval is only $\overline{\Pi_{>}^{\mathrm{vac}}}\ell=\tfrac{\epsilon_{0}}{2}n\ell$,
see table~\ref{tab:Densities}. The momentum surplus is the mechanical
momentum $\overline{\Pi^{\mathrm{mech}}}\ell,$ see table~\ref{tab:Densities}.
The mechanical momentum density can also be calculated from the Lorentz
force acting on the polarization current in the layers, at the border
of a signal, the result is the same. Instead of Eq.~(\ref{eq:p_Minkowski})
the relation between total signal momenta in medium and vacuum is
$p'=p\tfrac{1}{2}\left(n+1/n\right)$.  It can be checked with the
results in table~\ref{tab:Densities} that the center of energy of
a finite signal is conserved and that the component of the momentum
in the direction parallel to the surface is different outside and
inside the medium. The condition of no momentum transfer parallel
to the surface, used in a formal way in the isotropic case above,
should therefore be checked and justified.

Overall, the wave guide model reproduces the signal speed and law
of refraction of the corresponding continuum model, but the details
differ. The polarization tensor of the corresponding three dimensional
continuum model would be diagonal with components $\left(\epsilon_{0},\epsilon_{0}n^{2},\epsilon_{0}\right)$
or $\left(\epsilon_{0}n^{2},\epsilon_{0}n^{2},\epsilon_{0}\right)$,
which makes no difference for the polarization in $y$-direction considered
here. The continuum model does not give the correct total and mechanical
momentum of a light signal for the metamaterial of Fig.~\ref{fig:MetaMat}.

\section{Conclusions}

Wave packets in dispersion-free media move in the direction of their
momentum and obey momentum conservation like particles. This allows
to describe refraction of light on a quasi mechanical level. In any
case a substantial fraction of signal momentum is of mechanical nature,
and it is important to understand the corresponding internal structure
of wave packets.

Taking into account that atoms are accelerated but essentially remain
in place after the signal has passed leads to an electromagnetic system
with static matter. Despite this simplification, the problem remains
full of pitfalls and complications, related to the Abraham-Minkowski
debate. It is clear that most of the difficulties are due to limitations
of the usual effective medium Maxwell equations~\cite{Brevik1979,Silveir2017}.
These equations are adequate for effects determined by average polarization
and magnetization (linear in the fields), but not for bilinear local
terms like mechanical momentum, at least not without further input. 

We have used the exact solution for an anisotropic wave guide to demonstrate
these limitations. We do not claim that the well known results for
isotropic media are wrong, but what remains lacking is a better understanding
of the connection between microscopic physics and the conventional
effective medium theory. One could also speak of a complementarity
between microscopic and effective medium theories~\cite{Brevik1979},
but the relation between the formulations has not been examined in
detail. 

\section{Statements and declarations}

No datasets were generated or analyzed during the current study. The
author has no relevant financial or non-financial interests to disclose.

\medskip{}

\bibliographystyle{plain}
\bibliography{Other}

\end{document}